\documentclass[aps,pra,groupedaddress,showpacs]{revtex4}

\usepackage{graphicx}
\usepackage{amsmath}

\begin{document}


\title{Nonlinear dynamics for vortex lattice 
formation in a rotating Bose-Einstein condensate}


\author{Kenichi Kasamatsu}
\author{Makoto Tsubota}
\affiliation{Department of Physics,
Osaka City University, Sumiyoshi-Ku, Osaka 558-8585, Japan }
\author{Masahito Ueda}
\affiliation{
Department of Physics, Tokyo Institute of Technology,  
Meguro-ku, Tokyo 152-8551, Japan}

\date{\today}

\begin{abstract}
We study the response of a trapped Bose-Einstein condensate to a sudden turn-on of a 
rotating drive by solving the two-dimensional Gross-Pitaevskii equation. A weakly anisotropic 
rotating potential excites a quadrupole shape oscillation and its time evolution 
is analyzed by the quasiparticle projection method. A simple recurrence oscillation of 
surface mode populations is broken in the quadrupole resonance region that depends 
on the trap anisotropy, causing stochastization of the dynamics. 
In the presence of the phenomenological dissipation, an initially irrotational condensate
is found to undergo damped elliptic deformation followed by unstable surface ripple 
excitations, some of which develop into quantized vortices that eventually form a lattice. 
Recent experimental results on the vortex nucleation should be explained not only by 
the dynamical instability but also by the Landau instability; the latter is 
necessary for the vortices to penetrate into the condensate. 
\end{abstract}


\maketitle

\section{INTRODUCTION}{\label{intro}}
Since realization of a Bose-Einstein condensate (BEC) of alkali-metal atomic gases, 
much attention has been focused on the dynamical phenomena associated with superfluidity. 
The remarkable feature reflecting superfluidity appears in the response to external rotation. 
Recent observation of a quantized vortex lattice in trapped BECs 
\cite{Madison,Abo,Hodby,Haljan} confirmed the evidence of superfluidity. 
Madison {\it et al}. observed directly the nonlinear dynamical phenomena 
such as the vortex nucleation and lattice formation in the rotating condensate \cite{Madison2}. 
Such visualized results have greatly contributed to 
elucidation of static and dynamic properties of quantized vortices. 

The dynamics of dilute BECs has been successfully described by 
the Gross-Pitaevskii (GP) mean field model. 
For the quantized vortices in a trapped BEC, 
various theoretical studies have been made based on this model \cite{Fetterrev}. 
The mechanism of vortex nucleation in rotating trapped BECs is one of the important topics. 
Vortex nucleation of this system differs from that of a superfluid helium system in 
the ratio of the coherence length $\xi$ to the system size $L$. 
In the former where $\xi \alt L$, vortex nucleation is related 
to the instability of collective excitations whose energy scale is
set by the confining potential. In the latter where $\xi \ll L$, 
it is determined by the local dynamics.
A number of theoretical papers have discussed possible mechanisms of vortex 
nucleation 
\cite{Dalfovo,Isoshima,Feder1,Feder3,Ripoll,Sinha,Aftalion,Tsubota,Anglin,Muryshev,Mizushima,Simula,Williams,Kramer,Anglin2}. 
However, only a few papers made full numerical 
analysis of the time-dependent GP equation, which is necessary to 
understand the results in Ref. \cite{Madison2}. Although the imaginary time 
propagation of the GP equation is a powerful scheme to find equilibrium 
states \cite{Feder1,Aftalion}, the dynamical process toward such states 
cannot be revealed by this method. 
Feder {\it et al}. solved numerically the time-dependent GP 
equation in a rotating frame, but the motion of generated 
vortices remains turbulent, forming no vortex lattice 
\cite{Feder3}. Tsubota {\it et al.} \cite{Tsubota} included a phenomenological 
dissipation into the GP equation to simulate a vortex lattice formation, 
obtaining the results consistent with those of Madison {\it et al} \cite{Madison2}. 
The whole story of the vortex lattice formation has been clarified as follows:
(i) the condensate makes elliptic deformation to oscillate, 
(ii) surface waves are excited at the boundary of the condensate, 
(iii) quantized vortices enter the condensate from the boundary, forming a lattice. 

It has been not yet clear what relation between the dynamical 
processes (i)-(iii) and the intrinsic instability of a rotating condensate. 
There are two considerable instabilities for vortex nucleation, namely, 
the dynamical instability \cite{Sinha} and the Landau instability 
\cite{Dalfovo,Feder1,Anglin,Muryshev,Simula,Williams}. 
The former originates from the imaginary frequency of the excitation mode, 
giving rise to the exponential growth of the unstable mode even in the energy-conserving dynamics. 
The latter occurs when the corresponding excitation frequency becomes negative 
in the reference frame and some dissipation works. 
These two instabilities work generally in different parameter ranges. 
Thus, one may question which instability is important for the actual experiments on vortex nucleation. 
It should be noted that the vortex nucleation frequency in a series of ENS experiments 
\cite{Madison,Madison2} coincides with the prediction based on the dynamical instability \cite{Sinha}, 
contrary to some analyses based on the Landau instability. 

In this paper, we investigate theoretically the detailed 
dynamics of a BEC subject to external rotation by the numerical 
analysis of the two-dimensional GP equation, and solve the above questions. 
The first issue of this paper deals with a response of a BEC to a sudden 
turn-on of a rotating drive within the energy-conserving dynamics. 
The rotating potential excites chiefly the quadrupole surface 
mode with angular momentum $l=2$ and distorts the 
condensate into an ellipse. Because of the anisotropy of the trapping potential and 
the nonlinear atomic interaction, the different surface modes of oscillation are 
coupled to each other, causing complicated nonlinear dynamics. 
To clarify the mode coupling we use the quasiparticle projection 
method \cite{Morgan}, which allows us to decompose the 
macroscopic wave function into the condensate and noncondensate 
parts and determine the populations of each mode. 
We find that the condensate makes the simple periodic oscillation 
for most values of the rotation frequencies; populations in excited modes are recovered 
completely to initial values in the sense of the 
Fermi-Pasta-Ulam recurrence \cite{Fermi}. 
In resonance of the quadrupole mode, however, the simple 
recurrence is broken into chaotic behavior, 
reflecting the dynamical instability of the rotating condensate \cite{Sinha}. 
The increase of the trap anisotropy extends the range of the rotation 
frequency for the resonance excitation. 
The chaotic dynamics yields violent density and phase 
fluctuations at the condensate surface. 
The generated surface ripples slightly increase of the total angular momentum, 
but they never develop into quantized vortices in the energy-conserving simulation. 

Next, we consider the dissipative dynamics of the rotating BEC, 
extending our previous work \cite{Tsubota}. 
This paper describes more detailed dynamics 
of the vortex lattice formation by following the time development 
of the condensate density and phase simultaneously. 
The GP equation with phenomenological dissipation explains the 
experimental results very well. The quasiparticle projection method 
is also available in the analysis of the dissipative dynamics, 
thus we can study what modes are excited during the 
dynamical process of vortex lattice formation. 
In the presence of dissipation, the generation of surface ripple is caused by the 
surface modes with negative frequencies in the rotating frame \cite{Isoshima}; 
the onset of this instability is given by 
the Landau criterion applied to the rotating BEC \cite{Dalfovo,Feder1,Anglin,Muryshev}. 
Numerical simulation shows that the surface ripples induced by the Landau instability 
certainly develops to quantized vortices. Therefore, it is concluded that vortex nucleation 
is essentially caused by the Landau instability. 
The ENS experiments \cite{Madison} should be explained by the two-stage process: 
vortex nucleation by the Landau instability 
after the thermal component creation by the dynamical instability. 

This paper is organized as follows. Section \ref{model} introduces 
our formulation that describes the dynamics of a rotating BEC 
in a harmonic trap. In Sec. \ref{consedy}, we study the 
energy-conserving dynamics of a rotating BEC. The dynamics of 
the mode-coupling is analyzed by the quasi-particle projection method.
In Sec. \ref{lattice} we study the dissipative dynamics 
of the vortex generation and lattice formation in detail, 
and make some comments on the origin of the dissipation. 
Our results are compared with the experimental ones. 
Section \ref{conclusion} is devoted to the conclusion.

\section{THE MODEL} {\label{model}}
\subsection{Formulation of the problem}
A BEC trapped in an external potential is described by a 
``macroscopic wave function" $\Psi({\bf r},t)$ obeying 
the GP equation. In the frame rotating with the frequency $\Omega$
around the $z$ axis the GP equation reads 
\begin{equation}
i \hbar \frac{\partial \Psi}{\partial t} = 
\biggl( - \frac{\hbar^{2}}{2 m} \nabla^{2} + V_{\rm trap} 
+ V_{\rm rot} - \mu + g|\Psi|^{2}- \Omega L_{z} \biggr) 
\Psi. \label{gpe1}
\end{equation}
Here $g=4 \pi \hbar^{2} a/m$ represents the strength of interactions 
characterized by the $s$ wave scattering length $a>0$, 
$\mu$ the chemical potential, 
$L_{z}=-i \hbar (x \partial_{y} - y \partial_{x})$ 
the angular momentum. The wave function is normalized by the 
total particle number $N$ as $\int d{\bf r} |\Psi|^{2} =N$. 
An external harmonic trapping potential has the form
\begin{equation}
V_{\rm trap}({\bf r})=\frac{1}{2}m \{ \omega_{\perp}^{2} 
(x^{2} + y^{2}) + \omega_{z}^{2} z^{2} \},
\label{trappot}
\end{equation}
and a rotating potential 
\begin{equation}
V_{\rm rot}({\bf r})=\frac{1}{2}m \omega_{\perp}^{2} 
(\epsilon_{x} x^{2} + \epsilon_{y} y^{2}) 
\label{stirpot}
\end{equation}
with the anisotropy parameters $\epsilon_{x} \neq \epsilon_{y}$; 
this form describes approximately the rotating potential 
used in the ENS experiments \cite{Madison,Madison2}. 
Such a rotating potential breaks the rotational symmetry, 
thus transferring the angular momentum into the 
condensate through the excitation of surface modes or the generation of vortices. 
 
In order to reduce the system into the two-dimensional 
$x-y$ space, we separate the degrees of freedom of the 
wave function as 
$\Psi({\bf r},t)=\psi(x,y,t)\phi(z)$, obtaining 
the two-dimensional GP equation
\begin{eqnarray}
i \hbar \frac{\partial \psi(x,y,t)}{\partial t} = 
\biggl[ - \frac{\hbar^{2}}{2 m} 
\left( \frac{\partial^{2}}{\partial x^{2}} 
+ \frac{\partial^{2}}{\partial y^{2}} \right) \nonumber \\
+ \frac{1}{2} m \omega_{\perp}^{2} \{ (1+\epsilon_{x}) x^{2}
+(1+\epsilon_{y}) y^{2} \} \nonumber \\
- \mu + g \eta |\psi(x,y,t)|^{2} 
- \Omega L_{z} \biggr] \psi(x,y,t),
\label{2Dgpenama}
\end{eqnarray}
where 
\begin{equation}
\eta \equiv \frac{\int dz |\phi(z)|^{4}}{\int dz |\phi(z)|^{2}},
\end{equation}
and $\mu$ includes a constant arising from the integral 
of $\phi(z)$. The normalization of the two-dimensional wave 
function $\psi(x,y)$ is taken by the particle number 
$N_{\rm 2D}$ per unit length along the $z$-axis as
\begin{equation}
\int\int dx dy |\psi(x,y)|^{2} = 
N \left(\int dz |\phi(z)|^{2} \right)^{-1} = N_{\rm 2D}.
\end{equation}
It is convenient to introduce the scales characterizing the 
trapping potential; the length, time, wave function 
are scaled as
\begin{eqnarray}
x=a_{h}\tilde{x},\hspace{2mm} t = 
\frac{\tilde{t}}{\omega_{\perp}},\hspace{2mm} \psi
= \sqrt{N_{\rm 2D}}\frac{\tilde{\psi}}{a_{h}},\nonumber 
\end{eqnarray}
respectively, with $a_{h}=\sqrt{\hbar/2m\omega_{\perp}}$. 
Then the GP equation is reduced to a dimensionless form as 
\begin{eqnarray}
i \frac{\partial \psi}{\partial t} = 
\biggl[ - \left( \frac{\partial^{2}}{\partial x^{2}} 
+ \frac{\partial^{2}}{\partial y^{2}} \right) 
+ \frac{1}{4} \{(1+\epsilon_{x}) x^{2} \nonumber \\
+(1+\epsilon_{y}) y^{2} \} - \mu + C |\psi|^{2} 
- \Omega L_{z} \biggr] \psi,
\label{nondimgp}
\end{eqnarray}
where $C=8 \pi a \eta N_{\rm 2D}$ and the tilde is omitted for simplicity.

The two-dimensional approximation may be valid for the 
condensate in a ``pancake-shaped" potential 
($\lambda = \omega_{z}/\omega_{\perp} \gg 1$) or the 
central part of the condensate in a ``cigar-shaped" 
potential ($\lambda < 1$). These two types of situations yield 
different forms of the mean field interaction strength $C$. 
For $\lambda \gg 1$ and $\hbar \omega_{z}$ larger than 
the interaction energy, $\phi(z)$ is approximated 
by the one particle ground state wave function 
in a harmonic potential:
\begin{equation}
\phi(z) = \biggl( \frac{1}{\sqrt{2\pi} a_{hz}} \biggr)^{1/2} 
\exp \biggl(-\frac{z^{2}}{4a_{hz}^{2}} \biggr)
\end{equation}
with $a_{hz} = \sqrt{\hbar/2m\omega_{z}} = a_{h}/\sqrt{\lambda}$. 
Then, $N_{\rm 2D}=N$ and the parameter $C$ becomes
\begin{equation}
C= 8 \pi a N \eta = 4 \sqrt{\pi \lambda} N \frac{a}{a_{h}}.
\end{equation}
On the other hand, for a cigar-shaped condensate with $\lambda < 1$ 
one can approximate the system with cylindrical configuration, 
i.e., translation symmetry along the $z$ direction. 
By neglecting the spatial derivative term of $z$ component 
in Eq. (\ref{gpe1}), and the third term in Eq. (\ref{trappot}), 
the two-dimensional GP equation of Eq. (\ref{2Dgpenama}) 
is obtained. Then the parameter $C$ is written by 
\begin{equation}
C=8 \pi a \eta N_{\rm 2D} = 8 \pi a N \frac{\int |\phi(z)|^{4} dz}
{(\int |\phi(z)|^{2} dz)^{2}} \simeq \frac{8 \pi a N}{2R_{z}}.
\label{interastre}
\end{equation}
Here $R_{z}$ is assumed to be the Thomas-Fermi radius 
$\sqrt{2 \mu/m \omega_{z}^{2}}$ along the $z$ axis with 
the chemical potential evaluated at the parameter 
$\epsilon_{x,y} = 0$ and $\Omega = 0$: 
\begin{equation}
\mu = \hbar \omega_{\perp} 
\left( \frac{15}{8} N \lambda \frac{a}{a_{h}} \right)^{2/5}.
\end{equation}
The latter approximation is suitable for the ENS experiment 
\cite{Madison} which was made under the cigar-shaped potential, 
where $C$ (Eq. (\ref{interastre})) takes the value between 
$200$ and $500$. For the large condensate in the MIT 
experiment \cite{Abo} $C \sim 10000$, 
though they did not use the cigar-shaped potential.

In the two-dimensional analysis, the effect of vortex bending \cite{Ripoll2,Aftalion2}
is not taken into account. 
Recent experiments \cite{Rosenbusch} showed that the time scale of the 
vortex bending is found to be longer than $\geq$ 1 sec, 
which is much longer than the time scale of the dynamics of 
vortex lattice formation ($\sim$ 100 msec). 
We therefore consider our two-dimensional analysis effective for the present problem.

\subsection{Numerical scheme}
The numerical calculations 
of Eq. (\ref{nondimgp}) are done using an 
alternating direction implicit (ADI) method \cite{Cman}. 
Defining a time step 
$\delta_{t}$ and space meshes $\delta_{x} = \delta_{y} = \delta$, 
and denoting the discrete wave function as 
$\psi_{j,k}^{n}=\psi(j\delta_{x}, k\delta_{y}, n\delta_{t})$ 
, $\psi_{j,k}^{n}$ develops into $\psi_{j,k}^{n+1}$ 
via the intermediate state $\psi_{j,k}^{n+\frac{1}{2}}$ as 
\begin{eqnarray}
\psi_{j,k}^{n+\frac{1}{2}} = \psi_{j,k}^{n} 
- \frac{\Delta t}{\delta^{2}} 
(\partial_{x}^{2} \psi_{j,k}^{n+\frac{1}{2}} 
+ \partial_{y}^{2} \psi_{j,k}^{n}) \nonumber \\
-i \Omega \Delta t (k \partial_{x} \psi_{j,k}^{n+\frac{1}{2}} 
- j \partial_{y} \psi_{j,k}^{n}) \nonumber \\
+\frac{\Delta t}{2} \{ (V_{j,k} + C |\psi_{j,k}^{n+\frac{1}{2}}|^{2}) 
\psi_{j,k}^{n+\frac{1}{2}} \nonumber \\
+ (V_{j,k} + C |\psi_{j,k}^{n}|^{2}) \psi_{j,k}^{n} \}
\end{eqnarray}
and
\begin{eqnarray}
\psi_{j,k}^{n+1} = \psi_{j,k}^{n+\frac{1}{2}} 
- \frac{\Delta t}{\delta^{2}} 
(\partial_{x}^{2} \psi_{j,k}^{n+\frac{1}{2}} 
+ \partial_{y}^{2} \psi_{j,k}^{n+1}) \nonumber \\
-i \Omega \Delta t (k \partial_{x} \psi_{j,k}^{n+\frac{1}{2}} 
- j \partial_{y} \psi_{j,k}^{n+1}) \nonumber \\
+\frac{\Delta t}{2} \{ (V_{j,k} + C |\psi_{j,k}^{n+1}|^{2}) 
\psi_{j,k}^{n+1} \nonumber \\
+ (V_{j,k} + C |\psi_{j,k}^{n+\frac{1}{2}}|^{2}) 
\psi_{j,k}^{n+\frac{1}{2}} \}.
\end{eqnarray}
Here we denote $\Delta t \equiv \delta_{t}/2i$, 
$\partial_{x} \psi_{j,k}^{n} \equiv ( \psi_{j+1,k}^{n} 
- \psi_{j-1,k}^{n})/2 $, 
$\partial_{x}^{2} \psi_{j,k}^{n} \equiv \psi_{j+1,k}^{n} 
- 2 \psi_{j,k}^{n}+ \psi_{j-1,k}^{n}$ and 
$V_{j,k}=\{(1+\epsilon_{x}) (j \delta)^{2} 
+ (1+\epsilon_{y}) (k \delta)^{2}\}/4$. 
We used $[-128 \leq j,k \leq +128]$ discretized space for the 
two-dimensional numerical simulation. 
The time step $\delta_{t} = 1.0 \times 10^{-3}$ ensures 
the numerical stability over sufficiently long propagation. 

\section{QUADRUPOLE OSCILLATION OF A ROTATING BEC} \label{consedy}
\subsection{Time development of the deformation parameter} \label{consedy1}
We start by discussing the time evolution from the stationary solution in a non-rotating trap. 
We turn on a rotating drive following the experimental procedure 
of Madison {\it et al.} \cite{Madison2}. The rotation with
a frequency $\Omega$ starts at $t=0$, and the trap anisotoropy 
$\epsilon = \{ (1+\epsilon_{x}) - (1+\epsilon_{y}) \}/ 
\{ (1+\epsilon_{x}) + (1+\epsilon_{y}) \}$ is increased rapidly from zero 
to its final value 0.025 in 20 msec. The strength of interaction 
$C$ is set to be $500$, corresponding to $a=5.77$nm, 
$N=3 \times 10^{5}$, $\omega_{z}=11.8 \times 2 \pi$, 
$\lambda=\omega_{\perp}/\omega_{z}=9.2$ \cite{Madison2}. 
The unit for length is 
$a_{h}=\sqrt{\hbar/2 m \omega_{\perp}}=0.728 \mu$m and the 
period of the trap 9.21 msec.

Rapid modulation of the trapping anisotropy induces the elliptic oscillation of 
the condensate. The elliptic oscillation is characterized by the deformation 
parameter \cite{Recati,Madison2}
\begin{equation}
\alpha=-\Omega \frac{<\!x^{2}\!>-<\!y^{2}\!>}{<\!x^{2}\!>+<\!y^{2}\!>},
\label{deforma}
\end{equation}
where $<A>$ means $\int dxdy \psi^{\ast} A \psi$. The time evolution of $\alpha$ 
for several values of $\Omega$ is shown in Fig. \ref{oscicon}. 
For relatively small values of $\Omega$, $\alpha$ makes a simple
periodic oscillation with positive values; 
the initial axisymmetric condensate is elongated 
along the $y$ axis because of the small trap 
anisotropy $\epsilon$ ($\epsilon_{x}>\epsilon_{y}$).
As $\Omega$ increases from zero, both the amplitude and the period increase gradually, 
leading to the large amplitude oscillation near $\Omega=0.7\omega_{\perp}$. 
For $\Omega=0.75\omega_{\perp}$, however, the periodicity of the oscillation 
is broken as shown in Fig. \ref{oscicon}. 
As $\Omega$ increases further, the periodicity is again recovered, 
but the sign of $\alpha$ changes to negative and its absolute value 
become smaller and smaller. The negative $\alpha$ means 
that the longer axis of the condensate ellipse is perpendicular to 
the longer axis of the trapping anisotropy. 

This shape oscillation mainly consists of the collective 
surface mode with angular momentum $l=2$, i.e., quadrupole mode. 
In the Thomas-Fermi limit, the dispersion relation for the surface mode reduces to 
$\omega_{l}=\sqrt{l}\omega_{\perp}$ \cite{Stringari}. 
The centrifugal term $-\Omega L_{z}$ lifts the surface mode frequency just as $-l\Omega$. 
For $l=2$, hence, it is expected that the quadrupole mode is resonantly 
excited at $\Omega/\omega_{\perp}=\sqrt{2}/2 \simeq 0.707$. 
As discussed later, the complete resonance does not occur 
because the quadrupole mode couples with various higher-energy modes through 
the nonlinear interaction, giving rise to the complicated dynamics. 
We find that for the rotation range $0.72<\Omega/\omega_{\perp}<0.78$ 
the oscillation becomes irregular, reflecting the nonlinear character of the dynamics. 
The deviation from the pure resonance frequency 
$\Omega/\omega_{\perp}= 0.707$ is due to the effect of the trap anisotropy 
and the nonlinear interaction (see Eq. (\ref{quasiamptd})).

Figure \ref{dynvorinst} shows the profile of the condensate density $|\psi(x,y,t)|^2$ 
and the phase $\theta(x,y,t)=\tan^{-1} ({\rm Im} \psi/{\rm Re} \psi)$ ($0<\theta<2\pi$) 
when the irregular oscillation occurs. On the surface of the condensate, 
there appear surface ripples, that is violent density fluctuations with a short wave length. 
In addition, the phase profile shows that many phase singularities, i.e., 
quantized vortices, come into the condensate surface, 
while the phase keeps the form of the quadrupoler flow inside the condensate \cite{Feder1,Recati}.  
Since these singularities are on the outskirts of the condensate 
where the amplitude $|\psi|$ is very small, they hardly contribute 
to both the energy and the angular momentum. 
Such phase singularities, named ``ghost vortices" in our previous paper \cite{Tsubota}, 
are located in the density hollows produced by the surface ripples. 

In the energy-conserving dynamics, such an irregular dynamics is caused by 
the dynamical instability associated with the imaginary frequency of the excitation modes. 
Sinha and Castin \cite{Sinha} made a linear stability analysis of an oscillating condensate 
with a quadrupolar ansatz, and found the growth of the 
fluctuation written by polynomials of degree $n=3$ around $\Omega/\omega_{\perp}=0.73$. 
They proposed that the associated dynamical instability  
triggers vortex nucleation in a condensate rotated in an anisotropic harmonic 
potential, in excellent agreement with the experimental results of ENS \cite{Madison}. 
However, Fig. 2 suggests that the generated surface ripples should be described 
by polynomials with higher-order degrees than $n=3$. 
Thus, we also make a linear stability analysis for $n=4$ and 8 by following Ref. \cite{Sinha}, 
and find that the growth rate of these higher order modes is as much as $n=3$. 
Hence, the dynamical instability excites such 
higher order excitation mode, generating the surface ripples. 
These ripples slightly increase of the total angular momentum 
because of the presence of ghost vortices, but
they never penetrate the inside of the condensate to form a lattice. 
Therefore, it is concluded that the dynamical instability 
does not lead to ``vortex nucleation" as observed in the ENS experiments. 
In Sec. \ref{lattice}, we will show that the dissipation-assisted instability can 
make the surface ripples develop into the quantized vortices. 

\subsection{Quasiparticle projection method} \label{morganform} 
The dynamics of the elliptic oscillation is well understood by decomposing 
the whole dynamics into an assembly of fundamental excitation modes. 
The quasiparticle projection method, developed by Morgan 
{\it et al.} to study the nonlinear mixing of the collective excitations \cite{Morgan}, 
enables us to decompose a wave function into a condensate and noncondensate modes, 
and monitor the time evolution of their populations. 
Here we will use this method to study the time development 
of the surface modes excited by the anisotropic rotating trap. 

To construct the mode functions for the projection, we use the solution of the 
time-independent GP equation with the non-rotating 
axisymmetric trap
\begin{equation} 
\biggl[ - \frac{d^{2}}{d r^{2}} - \frac{1}{r} \frac{d}{d r} 
+ \frac{r^{2}}{4} + C |\psi_{g}|^{2} \biggr] \psi_{g}(r) 
= \mu \psi_{g}(r),
\end{equation}
where $\psi_{g}$ corresponds to the initial non-vortex state 
in our simulation. The quasiparticle mode functions 
$u_{i}({\bf r})=u_{i}(r) e^{i l \theta}$ and 
$v_{i}({\bf r})=v_{i}(r) e^{i l \theta}$ are obtained by 
the Bogoliubov-de Gennes equations 
\begin{subequations}
\begin{eqnarray}
\biggr[ -\biggl( \frac{d^{2}}{d r^{2}}+ \frac{1}{r} \frac{d}{d r} 
- \frac{l^{2}}{r^{2}} \biggr) + \frac{r^{2}}{4} - \mu 
+2 C |\psi_{g}|^{2} \biggl] u_{i}(r) \nonumber \\ 
+ C \psi_{g}^{2} v_{i}(r) = \omega_{i} u_{i}(r), \\ 
\biggr[-\biggl( \frac{d^{2}}{d r^{2}}+ \frac{1}{r} \frac{d}{d r} 
- \frac{l^{2}}{r^{2}} \biggr) + \frac{r^{2}}{4} - \mu 
+2 C |\psi_{g}|^{2} \biggl] v_{i}(r) \nonumber \\
 + C \psi_{g}^{\ast 2} u_{i}(r) = - \omega_{i} v_{i}(r). 
\end{eqnarray}
\label{BdGeq}
\end{subequations}
The mode functions are subject to the orthogonality 
and symmetry relations
\begin{subequations}
\begin{eqnarray}
\int d^{2}r \{ u_{i}({\bf r}) u_{j}^{\ast} ({\bf r}) 
- v_{i}({\bf r}) v_{j}^{\ast}({\bf r}) \} = \delta_{ij}, \\
\int d^{2}r \{ u_{i}({\bf r}) v_{j}^{\ast} ({\bf r}) 
- v_{i}({\bf r}) u_{j}^{\ast}({\bf r}) \} = 0. 
\end{eqnarray}
\label{uvnormal}
\end{subequations}
Following the method of Ref. \cite{Morgan}, we introduce a set 
of excitations which are orthogonal to $\psi_{g}$. 
This is achieved by projecting out the overlap with $\psi_{g}$ 
from the solutions of the Bogoliubov-de Genne equations, 
the modified quasiparticle wave functions being defined by
\begin{subequations}
\begin{eqnarray}
\tilde{u}_{i}({\bf r}) &=& u_{i}({\bf r})-c_{i}\psi_{g}({\bf r}) \\ 
\tilde{v}_{i}^{\ast}({\bf r}) & =&v_{i}^{\ast}({\bf r})
+c_{i}^{\ast}\psi_{g}({\bf r}),
\end{eqnarray}
\label{orthogouv}
\end{subequations}
where $c_{i}=\int d^{2}{\bf r}\left[\psi_g^* u_i\right]=
-\int d^{2}{\bf r}\left[\psi_g v_i\right]$. 
The orthogonal relations Eq. (\ref{uvnormal}) still 
hold for these modified wave functions. The wave function can be expanded as 
\begin{equation}
\psi({\bf r},t)= \{ 1+b_g(t) \} \psi_g({\bf r}) 
+\sum_{i>0}\left\{ \tilde{u}_i({\bf r})b_i(t) + 
\tilde{v}_i^*({\bf r})b_i^*(t)\right\}. 
\label{newdefinition}
\end{equation}
It is easy to show that $1+b_{g}$ and $b_{i}$  
satisfy the relations
\begin{eqnarray}
1+b_{g}(t) = \int d^2 r\hspace{0.5 em} \psi_{g}^{\ast}({\bf r}) 
\psi({\bf r},t) , \\ 
b_{i}(t) = \int d^2 r \left\{ \tilde{u}_{i}^{\ast}({\bf r}) 
\psi({\bf r},t) - \tilde{v}_{i}^{\ast}({\bf r}) 
\psi^{\ast}({\bf r},t) \right\}. 
\label{orthoproject}
\end{eqnarray}
The populations of the ground state and the excitation 
are given by $|1+b_g|^2$ and $|b_i|^2$. 

In the following discussion, we take only those mode functions that carry 
angular momentum $l$ but do not possess any radial node, i.e., surface mode. 
Here the index $i$ of $\tilde{u}_{i}$ and $\tilde{v}_{i}$ 
is replaced with $l$. 
For $l \neq 0$ the overlap integral $c_{l}$ vanishes because 
$\int_{0}^{2\pi} d \theta e^{il\theta}=0$, so that 
$\tilde{u}_{l}=u_{l}$ and $\tilde{v}_{l}=v_{l}$. 
Since our system has even-parity for the spatial coordinate, 
no surface modes with odd $l$ are excited.

\subsection{Fermi-Pasta-Ulam recurrence and chaotic dynamics} \label{mixingsurf}
By using Eq. (\ref{orthoproject}), we project the time evolution 
of the excitation modes $|b_{l}|^{2}$ with $l=2,4,\cdot\cdot\cdot,20$ from 
$\psi({\bf r},t)$. Figure \ref{quasiamp}(a) shows the time evolution 
of $|b_{2}|^{2}$, $|b_{4}|^{2}$, $|b_{6}|^{2}$ and 
$|1+b_{g}|^{2}$ for $\Omega = 0.7 \omega_{\perp}$; other 
$|b_{l}|^{2}$ are extremely small. 
The rotating potential of Eq. (\ref{stirpot}) excites most 
intensively the $l=2$ mode which causes the elliptic deformation. 
The small population of the higher modes ($l=4,6,\cdot\cdot\cdot$) 
also appear following the increase of $|b_{2}|^{2}$. 
Then populations of those excited modes return to the initial values 
almost completely. This simple recurrence repeats periodically in time. 

For $\Omega=0.75 \omega_{\perp}$ the evolution is somewhat complicated 
as illustrated in Fig. \ref{quasiamp}(b). 
The complete recurrence is broken, but some quasi-periodicity still remains. 
As compared with the simple recurrence dynamics, the depletion of the ground 
state population is remarkable, which means that 
various modes with higher energy are more excited. 
The superposition of such higher-energy modes produces 
the surface ripples in the condensate density as shown in Fig. \ref{dynvorinst}. 

In general points of view, we face the problem known as Fermi, Pasta and Ulam 
recurrence phenomenon \cite{Fermi}. 
They studied statistical behavior in the 
chain of nonlinearly coupled oscillators, and found a quasiperiodic behavior 
of this system characterized by returns of the energy to the initial excited mode. 
Later, it was shown that there exists a threshold for the onset of ``stochastization" 
which is brought by high-energy excitations. 
For BEC in an anisotropic potential, the nonlinearities inside the condensate may give 
rise to the stochasticity in its time evolution \cite{Kagan}, which has been found in 
the numerical simulation of the GP equation \cite{Sinatra,Villain}.

\subsection{Two-mode analysis}
The simple analysis of the equation of motion for the quasiparticle 
population helps us understand the behavior of 
Fig. \ref{quasiamp} (a) and (b). 
Substituting Eq. (\ref{newdefinition}) into Eq. (\ref{nondimgp}), 
we get
\begin{widetext}
\begin{eqnarray}
i \frac{\partial b_{g}}{\partial t} = \int d^{2}r \psi_{g}^{\ast} 
V_{\rm rot} (\psi_{g}+\Delta) + C \int d^{2} r 
\biggr[ (b_{g}+b_{g}^{\ast}) |\psi_{g}|^{4} + 2 |\Delta \psi_{g}|^{2} 
+ \Delta^{2} \psi_{g}^{\ast 2} + \Delta |\Delta|^{2} 
\psi_{g}^{\ast} \biggl],  \label{condamptd} \\
i \frac{\partial b_{l}}{\partial t} = (\omega_{l}-l\Omega) b_{l}(t) 
+ \int d^{2}r u_{l}^{\ast} V_{\rm rot} (\psi_{g}+\Delta) 
+ C \int d^{2} r u_{l}^{\ast} \biggr[ 2 |\Delta|^{2} \psi_{g} 
+ \Delta^{2} \psi_{g}^{\ast} + \Delta |\Delta|^{2} + |\psi_{g}|^{2} 
\psi_{g} (b_{g}+b_{g}^{\ast}) \biggl] \nonumber \\
+ \int d^{2}r v_{l}^{\ast} V_{\rm rot} 
(\psi_{g}+\Delta^{\ast}) 
+ C \int d^{2} r v_{l}^{\ast} \biggr[ 2 |\Delta|^{2} 
\psi_{g}^{\ast} + \Delta^{\ast 2} \psi_{g} + \Delta^{\ast} 
|\Delta|^{2} + |\psi_{g}|^{2} \psi_{g}^{\ast} (b_{g}+b_{g}^{\ast}) 
\biggl], \label{quasiamptd}
\end{eqnarray}
\end{widetext}
where $\Delta = b_{g} \psi_{g} + \sum_{l} \{ u_{l}({\bf r}) 
b_{l}(t) + v_{l}^{\ast}({\bf r}) b_{l}^{\ast}(t) \}$ and 
$V_{\rm rot} = (\epsilon_{x} x^{2} + \epsilon_{y} y^{2})/4$. 
The presence of $V_{\rm rot}$ makes the integrals 
$\int d^{2}r \psi_{g}^{\ast} V_{\rm rot} u_{l}$ or 
$\int d^{2}r u_{l'}^{\ast} V_{\rm rot} v_{l}$ $\cdot\cdot\cdot$ 
finite, and these terms are reduced to the form 
$T_{0}\delta_{l,l'}+T_{1}\delta_{l,l'-2}+T_{2}\delta_{l,l'+2}$ 
after the integral of $\theta$-component 
(if $\epsilon_{x}=\epsilon_{y}$, the off-diagonal terms vanishes). 
Hence, the ground state is coupled directly to only the $l=2$ mode in Eq. (\ref{condamptd}) 
through $V_{\rm rot}$. The remaining terms with $C$ couple the various modes with the 
same parity to each other, resulting in a complex time evolution.

As seen in Fig. \ref{quasiamp}(a), the dominant contribution to the dynamics 
comes from the change of the population $|1+b_{g}|^{2}$ and $|b_{2}|^{2}$. 
To understand the basic properties we take only terms with $l=0$ and 2 
(two-mode approximation). In addition, we neglect the contribution of $v_{2}(r)$, 
because this term represents the excitation traveling 
oppositely to the rotation. Then Eqs. (\ref{condamptd}) 
and (\ref{quasiamptd}) reduce to 
\begin{subequations} 
\begin{eqnarray} 
i\frac{\partial b_{g}}{\partial t} &=& T_{00} (1+b_{g}) 
+ P_{00} \{ 2 {\rm Re}(b_{g}) + |b_{g}|^{2} \} (1+b_{g}) \nonumber \\ 
&&+ P_{02} (1+b_{g}) |b_{2}|^{2} + T_{02} b_{2}, \label{grondpotime} \\ 
i\frac{\partial b_{2}}{\partial t} &=& (\omega_{2}-2\Omega + T_{22}) b_{2} 
+ P_{22} |b_{2}|^{2} b_{2} \nonumber \\ 
&&+ P_{02} \{ 2 {\rm Re}(b_{g}) + |b_{g}|^{2} \} b_{2} + T_{02} (1+b_{g}), 
\label{l2potime}
\end{eqnarray} 
\label{two-modeap}
\end{subequations}
where
\begin{eqnarray}
T_{00}\! &=& \! \int d^{2}r \psi_{g}^{\ast} V_{\rm rot} \psi_{g}, 
\hspace{3mm} T_{22} = \int d^{2}r u_{2}^{\ast} V_{\rm rot} u_{2} \nonumber \\ 
T_{02}\! &=& \! \int d^{2}r \psi_{g}^{\ast} V_{\rm rot} u_{2} = 
\int d^{2}r u_{2}^{\ast} V_{\rm rot} \psi_{g}, \nonumber \\ 
P_{00}\! &=& \! C \int d^{2}r |\psi_{g}|^{4}, \hspace{3mm}
P_{22} =  C \int d^{2}r |u_{2}|^{4}, \nonumber \\
P_{02} \! &=& \! 2 C \int d^{2}r |\psi_{g}|^{2} |u_{2}|^{2}. \nonumber
\end{eqnarray}
It is convenient to represent the complex values $b_{g}$ and $b_{2}$ in terms of 
the amplitude and the phase as $1+b_{g}=|1+b_{g}|e^{i\theta_{g}}$ 
and $b_{2}=|b_{2}|e^{i\theta_{2}}$. 
Since the total population $|1+b_{g}|^{2}+|b_{2}|^{2}$ is a constant of motion, 
Eqs. (\ref{two-modeap}) have two variables: population difference 
$p=|1+b_{g}|^{2}-|b_{2}|^{2}$ and relative phase $\theta=\theta_{g}-\theta_{2}$. 
Then, Eqs. (\ref{two-modeap}) reduce to 
\begin{eqnarray}
\frac{dp}{dt}=2T_{02}\sqrt{1-p^{2}} \sin \theta, \label{jos1} \\
\frac{d\theta}{dt}=U p - T_{02} \frac{2p}{\sqrt{1-p^{2}}} 
\cos \theta - \Delta \omega \label{jos2}
\end{eqnarray}
with the conserved Hamiltonian
\begin{equation}
H(p,\theta) = \frac{1}{2} U p^{2} + 2 T_{02} \sqrt{1-p^{2}} \cos \theta - \Delta \omega p,
\end{equation}
where $U = (P_{00}+P_{22}-2P_{02})/2$ and 
$\Delta \omega = \omega_{2} - 2 \Omega + T_{22} -T_{00} + U$. 
These formula are the same with those used in the Josephson dynamics of 
the condensate in a double well potential, and the exact solution of $p(t)$ 
is expressed by the Weierstrassian elliptic function \cite{Smerzi2}. 
For $C=500$, we numerically determine as $\omega_{2}=1.438 \omega_{\perp}$, $U=-0.05$, 
$\delta \omega=\omega_{2}-2 \Omega +3 \epsilon - 0.05$ and $T_{02}=1.14 \epsilon$ 
from the calculated $\psi_{g}$ and $u_{2}$. 
The solution of Eqs. (\ref{jos1}) and (\ref{jos2}) with these parameter 
values are shown in Figs. \ref{quasiamp}(c). 
The periodicity and the amplitude of $|1+b_{g}|^{2}$ reproduce well the results 
of the GP equation. 

Figure \ref{b2ampli} shows the dependence of the maximum of the $|b_{2}|^{2}$ 
oscillation on $\Omega/\omega_{\perp}$. 
The maximum grows near $\Omega=0.7\omega_{\perp}$. 
Note that as $\epsilon$ increases the peak is shifted to the 
larger value of $\Omega$ from the pure $l=2$ resonance frequency 
$\Omega=\omega_{2}/2=0.719 \omega_{\perp}$. The dynamics of the GP equation 
does not cause pronounced resonance of $|b_{2}|^{2}$ 
because the transition into the higher energy modes occurs. 
For $\epsilon=0.025$ we found that the chaotic oscillation occurs in $0.72< \Omega <0.78$, 
which corresponds to the range where the maximum of 
$|b_{2}|^{2}$ is more than 0.4 in Fig. \ref{b2ampli}. 
Following this criterion, we obtain the $\Omega-\epsilon$ parameter region 
where the large amplitude oscillation is expected, 
as shown in the inset of Fig. \ref{b2ampli}. 
In this region the mode coupling via the nonlinear interaction becomes important, 
leading to a chaotic dynamics. We confirm by the numerical simulation 
that the simple recurrence is indeed broken in this region. 
Such a nonlinear mode coupling seems to be important in the 
experiment on vortex nucleation \cite{Madison}, 
and the detailed will be discussed in Sec. \ref{surfaceex}. 

\section{DYNAMICS OF VORTEX LATTICE FORMATION} \label{lattice}
In this section we focus on the dissipative dynamics of a trapped condensate 
following the sudden turn-on of rotation. The dissipation is necessary to simulate the dynamics of 
vortex lattice formation by the GP equation, 
because a vortex lattice corresponds to a local minimum of the total energy 
in the configuration space \cite{rota,Tsubota2}. 
Preliminary results were reported in Ref. \cite{Tsubota}. 
This section deals with the detailed dynamics 
by following the time development of the condensate density and the phase simultaneously, 
and presents a number of previously unpublished results. 
Although the dissipation is treated phenomenologically in the GP equation, 
the simulation explains the experimental results very well. 

The results show that the generation of surface ripples is also 
achieved by the instability of negative eigenvalue modes, i.e., Landau instability, 
and the following time development certainly leads to vortex penetration into the condensate. 
The experimental results on vortex nucleation by the ENS group \cite{Madison,Madison2} 
can be understood by taking into account both the dynamical instability and the Landau instability. 
This is described in Sec. \ref{surfaceex}. 

\subsection{Phenomenological dissipative equation} 
Before discussing the detailed dynamics, we make some 
comments on the dissipation. 
As in the previous study \cite{Tsubota}, the dissipation is 
treated phenomenologically in the GP equation. 
The time derivative term of Eq. (\ref{nondimgp}) is modified as 
\begin{eqnarray}
(i-\gamma) \frac{\partial \psi}{\partial t} = 
\biggl[ - \left( \frac{\partial^{2}}{\partial x^{2}} 
+ \frac{\partial^{2}}{\partial y^{2}} \right) 
+ \frac{1}{4} \{(1+\epsilon_{x}) x^{2} \nonumber \\
+(1+\epsilon_{y}) y^{2} \} - \mu + C |\psi|^{2} 
- \Omega L_{z} \biggr] \psi,
\label{nondimdisgp}
\end{eqnarray}
where the dimensionless parameter $\gamma$ describes the dissipation. 
This form of the dissipative equation follows the study of Choi {\it et al.} \cite{Choi} 
and that of other superfluid systems \cite{Aranson}. 
Choi {\it et al.} determined the value of $\gamma$ to be 0.03 by
fitting their theoretical results with the MIT experiments on collective
damped oscillations \cite{Mewes}. 
Thus, we also use $\gamma =0.03$ throughout this work.
Since this dissipative term is much
smaller than other terms in the GP equation, a small variation of
$\gamma$ does not change the dynamics qualitatively but only modifies
the relaxation time scale. 

The phenomenological dissipative equation (\ref{nondimdisgp}) may be relevant to the recent 
numerical work by Jackson and Zaremba \cite{Jackson1} on the coupled dynamics 
of a condensate and a noncondensate. 
Their simulation is based on the the generalized GP equation at a finite temperature
\begin{equation}
i \hbar \frac{\partial \Psi}{\partial t} = 
\biggl( -\frac{\hbar^{2}\nabla^{2}}{2m} + 
V_{\rm trap} + g n +2 g \tilde{n} -i \Gamma \biggr) \Psi. \label{ZGNd}
\end{equation}
This equation was derived by Zaremba, Nikuni and Griffin \cite{Zaremba}, 
where $n({\bf r},t)=|\Psi({\bf r},t)|^{2}$ is the 
condensate density and $\tilde{n}({\bf r},t)$ the 
noncondensate density. The dynamics of the noncondensate 
was described by the Boltzmann kinetic equation for the 
distribution function of the noncondensate atoms. 
The noncondensate atoms are assumed to obey the single-particle 
Hartree-Fock spectrum in their formulation. The numerical simulation 
by Jackson and Zaremba \cite{Jackson1} well 
explains the experimental results by the JILA group \cite{Jin}. 

Equation (\ref{nondimdisgp}) can be derived from 
Eq. (\ref{ZGNd}) with some additional approximations. 
We treat the noncondensate as being in static thermal equilibrium, and neglect 
the mean field of the noncondensate under the assumption $n \gg \tilde{n}$. 
Then, the dissipation of the condensate 
motion is associated with the term 
$\Gamma({\bf r},t)=(\hbar/2n)\Gamma_{12} = 
(\hbar/2n)\int\{d{\bf p}/(2 \pi \hbar)^{3}\} C_{12}$ 
with the collision integral $C_{12}$ between the condensate 
and noncondensate atoms. Under the local equilibrium distribution 
of the thermal atoms, $\Gamma_{12}$ is proportional to the 
difference of the local chemical potential between condensate 
and noncondenate as 
$\Gamma_{12} \propto \mu_{\rm nc}({\bf r},t) - \mu({\bf r},t)$ 
\cite{Zaremba}. Approximating $\mu({\bf r},t)\Psi \simeq -i \hbar (\partial \Psi/\partial t)$ 
\cite{tyukoku}, we obtain 
\begin{equation}
(i-\gamma) \hbar \frac{\partial \Psi}{\partial t} = 
\biggl( - \frac{\hbar^{2}}{2 m} \nabla^{2} + V_{\rm trap} 
+ g|\Psi|^{2} - i \gamma \mu_{\rm nc} \biggr) \Psi. \label{gpekakan}
\end{equation} 
Compared with Eq. (\ref{nondimdisgp}), the chemical potential of 
this equation is replaced by that of noncondensate $\mu_{\rm nc}$, 
yielding the imaginary term $i \gamma \mu_{\rm nc}$. 
Note that the relation $\mu=\mu_{\rm nc}$ 
is satisfied for the equilibrium condensate. 
If the space and time dependence of $\mu_{\rm nc}$ is neglected, 
this equation becomes Eq. (\ref{nondimdisgp}) 
by the transformation $\Psi \rightarrow \Psi e^{-i \mu_{\rm nc} t / \hbar}$ 
and $\mu_{\rm nc} \rightarrow \mu$. 

The time development of Eq. (\ref{nondimdisgp}) does not conserve the norm of the wave 
function as well as the energy. In our simulation, the chemical potential 
$\mu$ is adjusted at each time step in order to conserve the 
norm and to decrease the total energy monotonically.
Recently, Eq. (\ref{gpekakan}) was also derived by Gardiner {\it et al.} 
by another approach \cite{Gardiner}, extended to the simulation 
of vortex lattice formation from the rotating thermal cloud \cite{Penckwitt}. 
They made the numerical simulation with the fixed chemical potential, 
finding that the norm of $\Psi$ decreased or increased as the system evolved. 
In the actual experiment, such a loss of the number of condensate atoms is 
caused by the rotation-induced heating \cite{Hodby}, so that
the generated thermal component might affect the dissipative dynamics of the vortex generation. 
However, under the phenomenological model, the change of the norm and the time scale 
of the dynamics is only a few \% by the effect, whether the chemical potential is fixed or not. 
The detailed quantitative study of those processes is beyond the scope of this paper; 
our treatment for the chemical potential is adequate to describe the 
actual dissipative dynamics. 

The assumption of the static noncondensate may be applicable to the experimental 
condition in Ref. \cite{Madison,Madison2,Abo,Hodby}.
According to the estimation by Guery-Odelin \cite{Odelin}, 
the spin up time for the whole noncondensed atoms to 
catch up with the rotating trap is about 15 seconds 
in collisionless regime. Since the typical time for the 
vortex formation of the condensate is a few hundred milliseconds, 
the condensate motion is separated from the noncondensate 
one under the rotating perturbation. However, the dynamic coupling of the 
mean-field between condensate and noncondenate causes 
the condensate motion to be damped (known as Landau damping), 
which is not included in Eq. (\ref{nondimdisgp}). 
Such a damping is several times larger than that by the 
$C_{12}$-collision at very low temperature \cite{Jackson1,Williams2}. 
Thus an additional damping may be provided by 
the full coupled dynamics of a condensate and a noncondensate.

The value of $\gamma$ is estimated by 
following the formulation for a uniform Bose gas \cite{Nikuni}. 
The parameter $\gamma$, in which length and energy are scaled 
by $a_{h}$ and $\hbar \omega_{\perp}$, has the form
\begin{equation}
\gamma=16 \sqrt{2\pi} A n a^{3} \biggl( \frac{a_{h}}{a} \biggr) 
\sqrt{\frac{T_{C}}{T}} \sqrt{\frac{\hbar \omega_{\perp}}{k_{B} T}}.
\end{equation}
Here $A$ is a factor of order unity for $T>0.5T_{C}$ and approaches 
zero as $T \rightarrow 0$. Using the typical experimental 
parameters, for example, $T=0.5 T_{C}$, $T_{C} = 500$nK, 
$a = 5.5$nm, $n \sim 10^{14}$/cm$^{3}$ and 
$\omega_{\perp} = 100 \times 2 \pi$Hz, we obtain 
$\gamma \sim A \times 10^{-2}$ which is consistent with $0.03$ 
used in this paper. 

\subsection{Dynamics from a non-vortex state to a vortex state}
Using Eq. (\ref{nondimdisgp}), we discuss the dynamics of 
vortex lattice formation in more details. 
As in the previous section, a sudden switch-on of rotation is made 
for the condensate with $C=500$.
Figure \ref{density} shows the time development of the 
condensate density $|\psi(x,y,t)|^2$ for $\Omega/\omega_{\perp}=0.7$ \cite{HP}. 
Initially, the condensate makes a quadrupole oscillation, 
but the oscillation is damped due to the dissipation. 
After a few milliseconds, the boundary surface of the condensate becomes unstable, 
generating the surface ripples which propagate along the surface 
as shown in Fig. \ref{density}(c). The excitations are likely 
to occur on the surface whose curvature is low, i.e. parallel 
to the longer axis of the ellipse. Then the waves on the surface 
develop into the vortex cores in a very short time [Fig. 
\ref{density}(d) and \ref{density}(e)]. 
As is well known in the study 
of rotating superfluid helium \cite{rota,Tsubota2}, 
the rotating drive pulls vortices into the rotation axis, while repulsive interaction 
between vortices tends to push them apart; this competition yields 
a vortex lattice whose vortex density  depends on the rotation 
frequency. In the presence of dissipation, 
six vortices enter the condensate, 
eventually forming a vortex lattice. As the vortex lattice is 
being formed, the axisymmetry of the 
condensate shape is recovered. 

The rotating potential $V_{\rm rot}$ has even-parity with respect to the coordinate. 
Accordingly, the dynamics is symmetric in the $x$-$y$ plane, 
and the number of the generated vortices is always even. 
The even-parity forbids the dynamics in which the system 
develops an asymmetric steady state. To remove this restriction, 
we introduced an infinitesimal artificial perturbation with 
odd-parity in $V_{\rm trap}$. This perturbation allows the system to develop 
into an asymmetric steady state as shown in Fig. 
\ref{density}(h), the energy of which is lower than that 
in Fig. \ref{density}(g). It takes a few hundred msec for 
the transition from Fig \ref{density}(g) to \ref{density}(h). 

The corresponding time development of the phase of $\psi(x,y,t)$ is shown in Fig. \ref{phase}. 
As seen in the energy-conserving dynamics,
as soon as the rotation starts, the phase field inside the condensate takes 
the form of quadrupolar flow $\theta(x,y)=\alpha x y + {\rm const}$, and  
just outside the Thomas-Fermi boundary there appear ghost vortices; for example, 
Fig. \ref{phase}(b) shows about 20 vortices. 
Ghost vortices move toward the rotation axis, but their invasion 
into the condensate is prevented at the Thomas-Fermi boundary. 
However, as the surface ripples are generated, the ghost vortices 
start to penetrate the condensate. There takes place the selection 
of the defects to penetrate, because their further invasion costs 
energy and angular momentum. For example, Fig. \ref{phase} shows 
that six vortices enter the condensate 
and form a lattice, while other excessive 
vortices are repelled and escape to the outside. 

As seen from Fig. \ref{density}, vortex invasion is likely to occur on the 
surface parallel to the longer axis of the ellipse. 
This is simply understood by the velocity field of the elliptic 
condensate ${\bf v}=\nabla \alpha x y = (\alpha y, \alpha x)$ as seen 
in Fig. \ref{phase}(b). Near the condensate surface parallel 
to the longer axis of an ellipse, this velocity field has 
the same direction as the velocity field made by the ghost 
vortices. There the additive velocity field ${\bf v}$ works 
as the attractive force which pulls the ghost vortices into 
the condensate. While the velocities of the condensate and 
the ghost vortices have opposite direction near the surface parallel 
to a shorter axis, where the condensate dislikes 
the invasion of the ghost vortices. 
Therefore, the vortices enter the condensate from the surface 
parallel to the longer axis. 

Figure \ref{disto} shows the time evolution of the deformation parameter 
and that of the angular momentum per atom 
$\ell_z/\hbar=\int dx dy \psi^{\ast} (L_{z}/\hbar) \psi$ 
in our dynamics of Fig. \ref{density} and \ref{phase}. 
They very well reproduce the experimental results of 
Ref. \cite{Madison2}. Before 300 msec, both $\alpha$ and $\ell_z/\hbar$ make damped 
oscillations. When vortices enter the condensate, $\alpha$ 
falls abruptly to a value below 0.05 and $\ell_z/\hbar$ increases to 4 
reflecting the number of the generated vortices. 

The final equilibrium value of $\ell_{z}$ depends on the number 
of the vortices which form a lattice.
Figure \ref{vornumber} shows the dependence of the number 
of vortices on the frequency $\Omega/\omega_{\perp}$ and 
the angular momentum per atom $\ell_{z}/\hbar$ for 
$C=$250, 500 and 1500 at 800 ms after the rotation starts. 
The increase in $C$ stabilizes the lattices of more vortices for the 
same frequency, and reduces the critical 
frequency at which the first vortex appears. 
Note that the value of $\ell_{z}/\hbar$ is about a half of 
the number of vortices. This is understood by the simple 
model in which the condensate with a vortex lattice makes a 
rigid-body rotation. Then the mean angular momentum per atom 
at $r=\sqrt{x^{2}+y^{2}}$ is $\ell_{z}/\hbar=m \Omega r^{2}$. 
The average of the angular momentum per atom averaged over 
the whole condensate is given by 
\begin{equation}
<\ell_{z}/\hbar> = \frac{\int |\psi|^{2} (\ell_{z}/\hbar) 
d {\bf r}}{\int |\psi|^{2} d {\bf r}}. 
\end{equation}
Assuming the spatially homogeneous density, 
we obtain $<\ell_{z}/\hbar> = m \Omega R^{2}/2 \hbar$ with 
the typical radius $R$ of the condensate. 
In the limit of a rigid-body rotation, the number of vortices $N_{\rm v}^{\rm lattice}$ 
at the rotation frequency $\Omega$ is given by Feynman's rule 
\begin{equation}
N_{\rm v}^{\rm lattice}=\pi R^{2} n_{\rm v}=\pi R^{2} \frac{2\Omega}{\kappa} 
= \frac{m \Omega R^{2}}{\hbar},
\end{equation}
where $n_{\rm v}$ represents the number of vortices per unit 
area and $\kappa$ the quanta of circulation. Hence, 
one obtains $<\ell_{z}/\hbar> = N_{\rm v}^{\rm lattice}/2 $. 
The numerical result better agrees with this estimation for larger 
$\Omega$ and larger $C$ because the condensate with a dense 
vortex lattice mimics a rigid-body rotation \cite{Feder2}. 
The small disagreement may be attributed to a small 
deviation from Feynman's rule and the inhomogeneous density. 

\subsection{Dynamics starting from an initial state 
with one vortex: metastable state}
Next, we discuss the time evolution starting from a 
one-vortex state. As seen from Fig. \ref{density} and 
\ref{phase}, as soon as the rotation is turned on to the 
irrotational condensate, the condensate makes the quadrupole 
deformation and its phase takes the form 
$\theta(x,y)=\alpha x y + {\rm const}$. 
This behavior will be changed if the dynamics starts from 
the initial state with one vortex which has already the 
circulating phase field. 
References \cite{Williams} and \cite{Kramer} discuss the stability of 
the condensate with a vortex for the quadrupole mode, and connect it
with the additional vortex formation. 
To investigate this problem, we prepare the initial state with 
one vortex for $C=500$ and $\Omega=0.4\omega_{\perp}$ larger than the 
thermodynamical critical frequency stabilizing one vortex 
state \cite{Feder1,Ripoll,Ripoll2,Isoshima} and start 
to rotate the system with $\Omega=0.7\omega_{\perp}$ as before. 
The time evolution of the phase is shown in Fig. \ref{phase1pon}. 
The numerical simulation reveals the 
nontrivial structure of the phase field; at the center of 
the condensate the phase maintains the circulation 
carried by an original vortex, while in the outer region the phase 
makes the quadrupolar flow [Fig. \ref{phase1pon}(b)]. 
Therefore, the condensate with one vortex also 
makes quadrupole deformation. The corresponding time evolution 
of the deformation parameter $\alpha$ 
is shown in Fig. \ref{disto} by the dotted line. 
The small amplitude of $\alpha$ compared to the previous result 
is due to the shift of the resonance frequency of the quadrupole 
mode because of the presence of a vortex \cite{Zambelli}. 

After that, the dynamics follows the same process as before. 
The final steady state is the lattice with seven vortices. 
This state is energetically higher than that 
of Fig. \ref{density}(h) with six vortices. Therefore, 
by starting from different initial states, one can obtain 
various metastable states with different configuration of a 
vortex lattice, as studied in rotating superfluid helium \cite{rota}. 
These metastable states are observable experimentally.

\subsection{Surface ripple excitation via the Landau instability}
As stated above, vortices are initially 
generated outside the Thomas-Fermi boundary of the condensate, 
where the energy cost to create defects is small because of 
the extremely low density. However, their penetration into 
the condensate was accomplished with the help of the surface 
ripples, induced by the instabilities in the non-vortex state. 
The quasiparticle projection method is useful to reveal the instability 
of the surface mode in this dissipative dynamics. 
Note that the time derivative term of Eq. (\ref{condamptd}) and (\ref{quasiamptd}) 
is modified as $(i-\gamma)\partial/\partial t$. 
Then, the non-vortex state becomes unstable when at least one 
excitation frequency $\tilde{\omega}_{l}=\omega_{l}-l\Omega+O(\epsilon)$ 
becomes negative, causing the exponential growth like 
$b_{l}(t) \sim e^{-\gamma \tilde{\omega}_{l} t}$. 
Isoshima and Machida examined that the instability associated with 
the negative excitation frequency gives rise to the vortex formation \cite{Isoshima}, 
and calculated the critical frequency $\Omega_{c}$ 
at which the first vortex appears within the Bogoliubov theory. 
Garc\'{i}a-Ripoll and P\'{e}rez-Garc\'{i}a calculated 
$\Omega_{c}$ for more realistic conditions \cite{Ripoll}. 
That critical frequency can be expressed by the Landau criterion 
applied to the rotating BEC \cite{Dalfovo}: 
\begin{equation}
\Omega_{c}={\rm min} \biggl( \frac{\omega_{l}}{l} \biggr).
\label{Lancri}
\end{equation}
The angular momentum $l_{c}$ which yields $\Omega_{c}$ takes 
a value more than 4 with the parameter used in experiments 
\cite{Isoshima,Dalfovo,Anglin}; for $C=500$ used in this paper, $l_{c}$=8 and 
$\Omega_{c}=0.5\omega_{\perp}$. 

Our simulation confirms that this instability actually 
leads to the vortex generation, where the hollows of the 
surface ripples always evolve into the vortex cores. 
The vortex core near the condensate surface has the size of the 
coherence length $\xi$ determined by the local density. 
The number of the vortices generated at the condensate surface may be  
approximately given by $N_{\rm v}^{\rm surf} \sim 2 \pi R / \xi$. 
The numerical solution shows $2 \pi R / \xi$ is nearly equal to $l_{c}$. 
This is understood by the fact that the surface mode 
with $l_{c}$ has the wavelength of the order of $\xi$ 
as discussed in Ref. \cite{Dalfovo}. 

Note that $N_{\rm v}^{\rm surf}$ differs generally from the 
number of vortices $N_{\rm v}^{\rm lattice}$, 
depending on $\Omega$, in an eventual 
vortex lattice. This fact classifies the dynamics of a 
vortex invasion into two regimes. 
When $N_{\rm v}^{\rm surf}>N_{\rm v}^{\rm lattice}$, the vortices 
which invade the condensate are chosen 
from $N_{\rm v}^{\rm surf}$ vortices generated at the surface
and form a lattice following the dissipative vortex dynamics, the extra
$N_{\rm v}^{\rm surf}-N_{\rm v}^{\rm lattice}$ vortices being expelled 
out \cite{Tsubota2}. 
This dynamics is shown in Fig. \ref{quasidisamp}(a) in terms 
of the quasiparticle populations for $N_{c} \simeq l_{c} = 8$ 
and $N_{\rm v}^{\rm lattice}=2$; for $\Omega=0.57\omega_{\perp}$ 
the excitation frequencies $\omega_{l}$ with $l=4 \sim 14$ are negative. 
At the moment the vortices are about to enter the condensate 
($t \sim 1.0$ sec), the modes with $l=$4, 6 and 8 
(with $|b_{l}|^{2}>0.02$) are excited. These modes rapidly 
decay as two vortices penetrate into the condensate. 
After the vortex invasion, the final growth of the projected population of
the $l=2$ mode reflects the phase structure 
in the surface region of the lattice with two vortices. 

On the other hand, if $N_{\rm v}^{\rm surf}<N_{\rm v}^{\rm lattice}$, 
the number of the vortices first generated at the condensate 
surface is not sufficient to form a final lattice, 
so that successive invasion of vortices is needed. 
This situation is shown in Fig. \ref{quasidisamp}(b) 
with $N_{\rm v}^{\rm lattice}=14$. 
The frequency $\tilde{\omega}_{2}$ is already negative at $\Omega=0.86\omega_{\perp}$, 
which causes the exponential growth of $|b_{2}|^{2}$.
When the condensate deforms elliptically at $t \sim 0.2$sec, many kinds of 
surface modes are excited violently. 
Then the quasiparticle populations with high angular momentum 
($l=10,12$ and 14 are plotted in Fig. \ref{quasidisamp}(b)) oscillates 
during a short time after this burst. 
Finally $|b_{14}|^{2}$ grows when 14 vortices 
enter the condensate completely. 
Note that the $l= 14$ mode does not grow shortly after the burst, 
but it grows gradually through the excitation of the lower angular momentum modes.
The observation of the time development 
of a density field could find that the radius of the condensate is 
increased unsteadily with the successive invasion of vortices.

\subsection{Relation between dynamical instability and Landau 
instability in the experiments on vortex nucleation} \label{surfaceex}
For an irrotational BEC subject to rotation, we have clarified by the 
numerical simulation of the GP equation that there exist two instabilities 
that is relevant for vortex generation. 
The dynamical instability appears when the quadrupole surface mode is resonantly excited. 
The Landau instability associated with the negative excitation 
frequency is effective only in the presence of the dissipation. 
In this section, we discuss which instability is important for actual vortex lattice formation
by referring to the experimental results. 

In the experiments \cite{Madison,Madison2,Abo}, the authors observed 
that the vortices are nucleated most near $\Omega=0.7\omega_{\perp}$. 
Thus, vortices are generated when the condensate becomes resonant 
with the rotating perturbation that excites a quadrupole mode. 
These results strongly support the dynamical instability 
scenario by Sinha and Castin \cite{Sinha}. 
However, they also observed the vortices at lower off-resonant frequency, 
which cannot understand by the dynamical instability.  
On the other hand, the critical frequency at which the first vortices 
appear is extensively discussed by several authors, based on the Landau criterion. 
\cite{Dalfovo,Isoshima,Ripoll,Anglin,Muryshev,Mizushima,Simula}. 
The relation between the dynamical instability and Landau instability is still controversial, 
thus we will discuss these relation in the range of on- and off-resonant frequency separately. 

In the on-resonant range $0.72<\Omega<0.78$, the surface ripple is excited, 
but the dissipation is necessary to generate vortices in a condensate. 
Here the dissipation originates in the thermal component, 
which should be almost negligible in the experiment of the atomic gas at very low temperatures. 
However, in the dynamical process of a condensate, there is a possibility that 
the thermal component will be produced under a strong perturbation. 
The experimental result of ENS \cite{Madison} may be explained as follows. 
Consider a situation in which there is almost no dissipation at very low temperatures, 
namely, Landau instability does not work there. In a quadrupole resonance, however, 
the dynamical instability causes stochasticity in condensate oscillations, 
leading to the creation of the thermal component \cite{Kagan,Sinatra,Villain}. 
Indeed, Hodby {\it et al.} in Oxford has reported that temperature of the system increases 
from $0.5T_{c}$ to $0.8T_{c}$ during the vortex formation procedure \cite{Hodby}. 
The time spent in this process is determined by the growth time of the dynamical instability, 
which is expected to be about 100 msec \cite{Sinha}.
Then, the created thermal component makes the dissipation effective, 
vortices penetrating into the condensate via the dissipation-assisted Landau instability. 
To make clear this hypothesis we need the analysis beyond the mean field 
and leave this issue for future study. 

In the off-resonant range, a condensate makes only a stable quadrupole oscillation without dissipation. 
However, our results show that vortices may be generated beyond $\Omega_{c}$ 
whenever finite dissipation works. 
Therefore, if we make the experiment in which the temperature is so high that the dissipation works effectively,
it is possible to observe the critical frequency given by Landau criterion.
For the weakly anisotropic rotating potential used in the ENS experiments and in this paper, 
excitation of high-$l$ modes can be made only through the $l=2$ excitation, so that 
it takes very long time for the vortex formation near the critical frequency.
Figure \ref{relaxatime} represents 
the relaxation time for vortex lattice formation in our simulation 
with $C=500$ and $\gamma=0.03$ ($\Omega_{c}=0.5$). 
Each relaxation time is taken at the moment when the angular momentum becomes a half of 
the final equilibrium value during the rapid increase (see Fig. \ref{disto}). 
We also show the growth time via Laudau instability 
$\tau_{l}=-\{\gamma(\omega_{l}-l\Omega)\omega_{\perp}\}^{-1}$ for $l=2,4$ and 6. 
For $\Omega<0.7\omega_{\perp}$, where $\omega_{2}-2\Omega>0$, 
the relaxation time is longer than $\tau_{l}$ with $l=4$, 6, 8,..., because high-$l$ mode is 
only excited through the $l=2$ mode which has no instability.
Near the critical frequency $\Omega_{c}$ one finds too long relaxation time about 1 sec. 
As $\Omega$ increases higher than 0.8$\omega_{\perp}$, where $\omega_{2}-2\Omega<0$, 
the relaxation time matches $\tau_{2}$. 
The ENS group \cite{Madison} did not continue the observation 
beyond 1 sec, finding no vortices in the off-resonant range near $\Omega_{c}$. 
The longer rotation and the presence of the dissipation could nucleate vortices 
in the rotation range beyond $\Omega_{c}$. 

\section{CONCLUSION}\label{conclusion}
We investigate the detailed dynamics of a rotating BEC 
in a trapped condensate following the sudden turn-on of rotation. 
The numerical analysis of the two-dimensional GP equation shows a series 
of nonlinear dynamics that has not been clarified so much. 
In the energy-conserving dynamics, we study the quadrupole oscillation 
induced by the anisotropic rotating potential, and the time development of 
excitation modes by the quasiparticle projection method. 
In the resonance range of the quadrupole mode, the large amplitude oscillation causes 
the nonlinear mode coupling towards the higher energy modes, reflecting the dynamical instability. 
This nonlinear process leads to generate surface ripples, but not to nucleate vortices.
In the dissipative dynamics, the vortex lattice formation is revealed in more detail. 
The vortex penetration into the condensate is achieved by the surface 
modes excitation with negative frequencies in the rotating frame, 
so that the critical frequency for vortex generation is determined by the Landau instability.
Two possible instabilities for vortex generation is discussed by comparison with the experiments.  
Although the dynamical instability may promote vortex formation, such an instability alone
cannot explain the experimental results. 
The dynamical instability plays the role of increasing a thermal component, 
and vortex nucleation and penetration is caused by the Landau instability. 

The similar situation with the competition between the dynamical instability and the 
Landau instability has been found in the center-of-mass oscillation of a BEC 
in a one-dimensional optical lattice \cite{Burger}. 
They observed the critical superfluid speed above which the dissipative dynamics starts. 
Its main features, including the parameter range of instability, can be explained 
by the simulation of the dissipationless GP equation \cite{Wu}, 
which means that the observed phenomena may be caused by the dynamical instability. 
However, the following development of a condensate is not explained only by 
the dynamical instability; the dissipation-assisted instability such as the Landau instability 
is necessary to describe the dynamics \cite{Burger}. 
These problems on the dissipative dynamics of a BEC, 
as well as the vortex lattice formation, offers the testing ground 
for the analysis beyond the framework of the GP equation. 


\begin{acknowledgments}
We would like to thank T. Nikuni for useful discussions.
We also acknowledge T. Iida for instructive comments on this work.
M.T. acknowledges support by a Grant-in-Aid for Scientific Research
(Grant No.12640357) by the Japan Society for the Promotion of Science.
M.U. acknowledges support
by a Grant-in-Aid for Scientific Research
(Grant No.11216204) by the Ministry of Education, Culture, Sports,
Science and Technology of Japan, and by the Toray Science Foundation.
\end{acknowledgments}


\begin{figure}[ht]
\caption{Time evolution of the deformation parameter $\alpha$ 
for $\Omega/\omega_{\perp}=$0.65, 0.70, 0.75, 0.80, 0.85.}
\label{oscicon}
\end{figure}

\begin{figure}
\caption{The density and phase profile of the condensate 
with $\Omega/\omega_{\perp}=0.75$ at 105 msec. 
In (b), the value of the phase changes continuously 
from $0$ (black) to $2 \pi$ (white).
There appear some lines where the phase changes discontinuously 
from black to white. These lines correspond to the branch 
cuts between the phases 0 and 2$\pi$, and their apexes around 
which the value of the phase rotates continuously from 0 to $2\pi$ 
represent phase defects, i.e., quantized vortices.}
\label{dynvorinst}
\end{figure}

\begin{figure}[ht]
\caption{Time evolution of the surface mode populations 
$|b_{l}|^{2}$ for (a) $\Omega/\omega_{\perp}=0.7$ and 
(b) $\Omega/\omega_{\perp}=0.75$. Figures (c) represents the time evolution 
of $|1+b_{g}|^{2}$ (dashed curve) and $|b_{2}|^{2}$ (solid curve) 
in the two-mode approximation}
\label{quasiamp}
\end{figure}

\begin{figure}[ht]
\caption{The dependence of the maximum of $|b_{2}|^{2}$ on $\Omega/\omega_{\perp}$ 
in the two-mode approximation 
for $\epsilon=$0.01, 0.025, 0.05 and 0.08. The inset shows the parameter 
region in which the large amplitude oscillation with max$(|b_{2}|^{2})>0.4$ occurs.} 
\label{b2ampli} 
\end{figure} 

\begin{figure*}
\caption{Time development of the condensate density $|\psi|^2$ 
after the trapping potential suddenly begins to rotate at $t=0$ with 
$\Omega=0.7\omega_{\perp}$.}
\label{density}
\end{figure*}

\begin{figure*}
\caption{Phase profiles of $\psi$ corresponding to the density profiles of 
Fig. \ref{density}. The value of the phase changes continuously 
from $0$ (black) to $2 \pi$ (white).
The discontinuous lines between black and white correspond 
to the branch cut of the complex plane, and their 
edges represent quantized vortices. 
The unit of length is the same as that of Fig. \ref{density}.}
\label{phase}
\end{figure*}

\begin{figure}[ht]
\caption{Time evolution of the distortion parameter $\alpha$ 
(solid curve) and the angular momentum per atom $\ell_{z}/\hbar$ 
(dashed curve) in the dynamics of Fig.\ref{density} and \ref{phase}. 
The dotted curve shows $\alpha$ for the dynamics starting 
from the initial state with one vortex.}
\label{disto}
\end{figure}

\begin{figure}[ht]
\caption{The number of vortices (solid curve) and angular momentum 
per atom (dotted curve) versus $\Omega/\omega_{\perp}$, 
for $C=$250 (with empty circles), 500 (with filled circles) 
and 1500 (with crosses).}
\label{vornumber}
\end{figure}

\begin{figure}[ht]
\caption{Phase profile of the simulation starting from 
one vortex state.}
\label{phase1pon}
\end{figure}

\begin{figure}[ht]
\caption{Time evolution of the surface-mode amplitude $|b_{l}|^{2}$ 
in the dissipative dynamics for (a) $\Omega=0.57\omega_{\perp}$ and 
(b) $\Omega=0.86\omega_{\perp}$ under the same condition in Sec. \ref{lattice}. 
In (b), we plot only modes with $l=2,10,12,14$, 
which are important in the discussion.
The inset shows the density profile of the final steady state. 
The other inset in (b) shows $|b_{10}|^{2}$, 
$|b_{12}|^{2}$ and $|b_{14}|^{2}$ near $t=0.3$ sec.}
\label{quasidisamp}
\end{figure}

\begin{figure}[ht]
\caption{The circles represent the relaxation time of vortex lattice formation 
by the numerical simulation for $C=500$ and $\gamma=0.03$. Thin solid curves show the $\tau_{l}=
-\{\gamma(\omega_{l}-l\Omega)\omega_{\perp}\}^{-1}$ for $l=2$, 4 and 6 obtained from Eq. (\ref{BdGeq})}
\label{relaxatime}
\end{figure}


\begin{thebibliography}{99}
\bibitem{Madison}
K. W. Madison, F. Chevy, W. Wohlleben, and J. Dalibard, 
Phys. Rev. Lett {\bf 84}, 806 (2000); 
F. Chevy, K. W. Madison, and J. Dalibard, 
{\it ibid}, {\bf 85}, 2223 (2000).
\bibitem{Abo}
J. R. Abo-Shaeer, C. Raman, J. M. Vogels, and W. Ketterle, 
Science, {\bf 292}, 476 (2001); 
C. Raman, J. R. Abo-Shaeer, J. M. Vogels, K. Xu, and W. Ketterle, 
Phys. Rev. Lett. {\bf 87}, 210402 (2001).
\bibitem{Hodby}
E. Hodby, G. Hechenblaikner, S. A. Hopkins, O. M. Marag\'{o}, 
and C. J. Foot,  Phys. Rev. Lett. {\bf 88}, 010405 (2002).
\bibitem{Haljan}
P. C. Haljan, I. Coddington, P. Engels, and 
E. A. Cornell, Phys. Rev. Lett. {\bf 87}, 210403 (2001). 
\bibitem{Madison2}
K. W. Madison, F. Chevy, V. Bretin, and J. Dalibard, 
Phys. Rev. Lett {\bf 86}, 4443 (2001).
\bibitem{Fetterrev}
For a review, see, for example, 
A. L. Fetter, and A. A. Svidzinsky, 
J. Phys. Condens. Matter {\bf 13}, R135 (2001). 
\bibitem{Dalfovo}
F. Dalfovo, S. Giorgini, M. Guilleumas, L. Pitaevskii, and S. Stringari, 
Phys. Rev. A {\bf 56}, 3840 (1997); 
F. Dalfovo and S. Stringari, Phys. Rev. A {\bf 63}, 011601 (2001).
\bibitem{Isoshima}
T. Isoshima and K. Machida, Phys. Rev. A {\bf 60}, 3313 (1999).
\bibitem{Feder1}
D. L. Feder, C. W. Clark, and B. I. Schneider, 
Phys. Rev. A {\bf 61}, 011601 (1999). 
\bibitem{Feder3}
D. L. Feder, A. A. Svidzinsky, A. L. Fetter, 
and C. W. Clark, Phys. Rev. Lett {\bf 86}, 564 (2001). 
\bibitem{Ripoll}
J. J. Garc\'{i}a-Ripoll and V. M. P\'{e}rez-Garc\'{i}a, 
Phys. Rev. A {\bf 63}, 041603(R) (2001). 
\bibitem{Sinha}
S. Sinha and Y. Castin, Phys. Rev. Lett. {\bf 87}, 190402 (2001). 
\bibitem{Aftalion}
A. Aftalion and Q. Du, Phys. Rev. A {\bf 64}, 063603 (2001). 
\bibitem{Tsubota}
M. Tsubota, K. Kasamatsu, and M. Ueda, 
Phys. Rev. A {\bf 65}, 023603 (2002). 
\bibitem{Anglin}
J. R. Anglin, Phys. Rev. Lett. {\bf 87}, 240401 (2001). 
\bibitem{Muryshev}
A. E. Muryshev and P. O. Fedichev, cond-mat/0106462. 
\bibitem{Mizushima}
T. Mizushima, T. Isoshima, and K. Machida, Phys. Rev. A {\bf 64}, 043610 (2001). 
\bibitem{Simula}
T. P. Simula, S. M. M. Virtanen, and M. M. Salomaa, Phys. Rev. A {\bf 66}, 035601 (2002).
\bibitem{Williams}
J. E. Williams, E. Zaremba, B. Jackson, T. Nikuni, 
and A. Griffin, Phys. Rev. Lett. {\bf 88}, 070401 (2002). 
\bibitem{Kramer}
M. Kraemer, L. Pitaevskii, S. Stringari, and F. Zambelli, 
Laser Physcis {\bf 12}, 113 (2002). 
\bibitem{Anglin2}
J. R. Anglin,  Phys. Rev. A {\bf 65}, 063611 (2002). 
\bibitem{Morgan}
S. A. Morgan, S. Choi, K.Burnett, and M. Edwards, 
Phys. Rev. A {\bf 57}, 3818 (1998). 
\bibitem{Fermi}
E. Fermi, J. Pasta, and S. Ulam, {\it Collected Papers of Enrico Fermi} 
(edited by E. Segr\'{e}, University of Chicago Press, Chicago, 1965). 
\bibitem{Ripoll2}
J. J. Garc\'{i}a-Ripoll and V. M. P\'{e}rez-Garc\'{i}a, 
Phys. Rev. A {\bf 64}, 053611 (2001). 
\bibitem{Aftalion2}
A. Aftalion and T. Riviere, Phys. Rev. A {\bf 64}, 043611 (2001). 
\bibitem{Rosenbusch}
P. Rosenbusch, V. Bretin, and J. Dalibard, cond-mat/0206511.
\bibitem{Cman}
W. H. Press et.al., {\it Numerical Recipes in C} 
(Cambridge University Press, Cambridge, England, 1988). 
\bibitem{Recati}
A. Recati, F. Zambelli, and S. Stringari, 
Phys. Rev. Lett {\bf 86}, 377 (2001). 
\bibitem{Stringari}
S. Stringari, Phys. Rev. Lett {\bf 77}, 2360 (1996).
\bibitem{Kagan}
Y. Kagan, E. L. Surkov, and G. V. Shlyapnikov, Phys. Rev. A {\bf 55}, R18 (1997). 
\bibitem{Sinatra}
A. Sinatra, P. O. Fedichev, Y. Castin, J. Dalibard, 
and G. V. Shlyapnikov, Phys. Rev. Lett. {\bf 82}, 251 (1999). 
\bibitem{Villain}
P. Villain and M. Lewenstein, Phys. Rev. A {\bf 62}, 043601 (2000). 
\bibitem{Smerzi2}
A. Smerzi, S. Fantoni, S. Giovanazzi, and S. R. Shenoy, 
Phys. Rev. Lett. {\bf 79}, 4950 (1997); S. Raghavan, A. Smerzi, 
S. Fantoni, and S. R. Shenoy, Phys. Rev. A {\bf 59}, 620 (1999).
\bibitem{rota}
L. J. Campbell and R. M. Ziff, Phys. Rev. B {\bf 20}, 1886 (1979).
\bibitem{Tsubota2}
M. Tsubota and H. Yoneda, J. Low Temp. Phys. {\bf 101}, 815 (1995).
\bibitem{Choi}
S. Choi, S. A. Morgan, and K. Burnett, Phys. Rev. A {\bf 57}, 4057 (1998); 
In this paper, the physical interpretation of the damping mechanism 
is the relaxation process of the thermal particles into the condensate. 
Thus, $\gamma$ represents the rate at which the excited 
components turn into the condensate. 
\bibitem{Aranson}
For example, I. Aranson and V. Steinberg, Phys. Rev. B {\bf 54}, 13072 (1996).
\bibitem{Mewes}
M. -O. Mewes, M. R. Andrews, N. J. van Druten, D. M. Kurn, 
D. S. Durfee, C. G. Townsend, and W. Ketterle, 
Phys. Rev. Lett. {\bf 77}, 988 (1996). 
\bibitem{Jackson1}
B. Jackson and E. Zaremba, Phys. Rev. Lett. {\bf 87}, 100404 (2001); 
{\bf 88}, 180402 (2002).
\bibitem{Zaremba}
E. Zaremba, T. Nikuni, and A. Griffin, 
J. Low Temp. Phys. {\bf 116}, 277 (1999). 
\bibitem{Jin}
D. S. Jin, M. R. Matthows, J. R. Ensher, C. E. Wieman, and E. A. Cornell,
Phys. Rev. Lett. {\bf 78}, 764 (1997).
\bibitem{Nikuni}
T. Nikuni, E. Zaremba, and A. Griffin, 
Phys. Rev. Lett. {\bf 83}, 10 (1999). 
\bibitem{tyukoku}
Substituting $\Psi({\bf r},t)=\sqrt{n({\bf r},t)} 
e^{i \theta({\bf r},t)}$ into the GP equation, we obtain 
\begin{eqnarray}
\frac{\partial n}{\partial t} = - \nabla (n {\bf v_{c}}), \\
\hbar \frac{\partial \theta}{\partial t} = \mu + \frac{1}{2}mv_{c}^{2},
\end{eqnarray}
where $\mu({\bf r},t)=(\hbar^{2} \nabla^{2} \sqrt{n})/
(2m\sqrt{n})+V_{\rm trap}+gn$ and ${\bf v_{c}}
=\hbar \nabla \theta/m$. Using these relations,
\begin{eqnarray}
-i \hbar \frac{\partial \Psi}{\partial t}&=&-i \hbar 
\frac{\partial \sqrt{n}}{\partial t} e^{i \theta} + 
\hbar \Psi \frac{\partial \theta}{\partial t} \nonumber \\
&=& \biggl(i\hbar\frac{\nabla(n{\bf v_{c}})}{2n}
+\frac{mv_{c}^{2}}{2}+\mu \biggr) \Psi.
\end{eqnarray}
If we assume $|{\bf v_{c}}|$ and the gradient of the density 
to be small, the approximation 
$\mu({\bf r},t)\Psi \simeq -i \hbar (\partial \Psi/\partial t)$ 
is attainable.
\bibitem{Gardiner}
C. W. Gardiner, J. R. Anglin, and T. I. A. Fudge, 
J. Phys. B {\bf 35}, 1555 (2002). 
\bibitem{Penckwitt}
A. A. Penkwitt, R. J. Ballagh, and C. W. Gardiner, cond-mat/0205037.
\bibitem{Odelin}
D. Guery-Odelin, Phys. Rev. A {\bf 62}, 033607 (2000). 
\bibitem{Williams2}
J. E. Williams and A. Griffin, Phys. Rev. A {\bf 63}, 023612 (2001); 
{\bf 64}, 013606 (2001). 
\bibitem{HP}
You can see the animation of this dynamics in
http://matter.sci.osaka-cu.ac.jp/bsr/vortexex-e.html. 
\bibitem{Feder2}
D. L. Feder and C. W. Clark,  Phys. Rev. Lett. {\bf 87}, 190401 (2001).
\bibitem{Zambelli}
F. Zambelli and S. Stringari, Phys. Rev. Lett. {\bf 81}, 1754 (1998). 
\bibitem{Burger}
S. Burger, F. S. Cataliotti, C. Fort, F. Minardi, M. Inguscio, 
M. L. Chiofalo, and M. P. Tosi, Phys. Rev. Lett. {\bf 86}, 4447 (2001);  {\bf 89}, 088902 (2002). 
\bibitem{Wu}
B. Wu and Q. Niu, Phys. Rev. Lett. {\bf 89}, 088901 (2002). 
\end{thebibliography}
\end{document}